\documentclass{elsart}
\usepackage{amssymb}

\begin{document}

\begin{frontmatter}

\title{The Invar Tensor Package}
\author[a]{J.M. Mart\'{\i}n-Garc\'{\i}a}%
\address[a]{Instituto de Estructura de la Materia, CSIC, \\
C/ Serrano 123, Madrid 28006, Spain}
\author[b]{R. Portugal}%
\author[b]{, and L.R.U. Manssur}%
\address[b]{Laborat\'orio Nacional de Computa\c{c}\~ao Cient\'{\i}fica (LNCC), \\
Av. Get\'ulio Vargas 333, Petr\'opolis, RJ, CEP 25651-075, Brazil}


\begin{abstract}
The {\em Invar} package is introduced, a fast manipulator of generic
scalar polynomial expressions formed from the Riemann tensor of a
four-dimensional metric-compatible connection. The package can maximally
simplify any polynomial containing tensor products of up to seven
Riemann tensors within seconds. It has been implemented both in
Mathematica and Maple algebraic systems.
\end{abstract}


\end{frontmatter}

{\footnotesize
\section*{Program summary}
\textit{Title of program:} Invar Tensor Package
\\
\textit{Catalogue identifier:}
\\
\textit{Program obtainable from:} CPC Program Library, Queen's
University of Belfast, N. Ireland
\\
\textit{Reference in CPC to previous version:} Computer Physics
Communications 157 (2004) 173-180
\\
\textit{Catalogue identifier of previous version:} ADSP
\\
\textit{Does the new version supersede the original program?:} No.
The original version runs only in Maple and its purpose is to be a
kernel for tensor manipulators. The present versions run in Maple
and Mathematica and are full tensor expression manipulators.
\\
\textit{Computers:} Any computer running Mathematica versions 5.0
to 5.2 or Maple versions 9 and 10
\\
\textit{Operating systems under which the new version has been
tested:} Linux, Unix, Windows XP
\\
\textit{Programming language:} Mathematica and Maple
\\
\textit{Memory required to execute with typical data:} 30 Mb
\\
\textit{No. of bits in a word:} 64 or 32
\\
\textit{No. of processors used:} 1
\\
\textit{No. of bytes in distributed program, including test data,
etc.:} Around 14 Mb
\\
\textit{Distribution format:} Uuencoded compressed tar file
\\
\textit{Nature of physical problem:} Manipulation and
simplification of tensor expressions. Special attention on
simplifying scalar polynomial expressions formed from the Riemann
tensor on a four-dimensional metric-compatible manifold.
\\
\textit{Method of solution:} Algorithms of computational group
theory to simplify expressions with tensors that obey permutation
symmetries. Tables of syzygies of the scalar invariants of the
Riemann tensor.
\\
\textit{Restrictions on the complexity of the problem:} The
present versions do not fully address the problem of reducing
differential invariants or monomials of the Riemann tensor with
free indices.
\\
\textit{Typical running time:} Less than a second to fully reduce
a monomial of the Riemann tensor of degree 7 in terms of
independent invariants.
}

\section{Introduction}
The Riemann tensor plays the most essential role in the geometrical
description of curved manifolds. Therefore it is a central object
of study in General Relativity and Cosmology, containing all invariant
(i.e. coordinate independent) and local information about the spacetime.
However, it is often very difficult to manipulate expressions containing
the Riemann tensor, due to its nontrivial algebraic properties.
In the last decade new mathematical ideas have drawn much renewed
attention to the subproblem of the algebraic scalars of the Riemann
tensors, but we are still far from having a simple and efficient
algorithm to manipulate generic expressions. Such an algorithm would
be very useful in several areas of Mathematics and Physics, for
example in the following particular cases: the generic manipulation of
Riemann-related tensors, like Lanczos-type potentials for the Weyl
tensor in General Relativity \cite{EdgarSenovilla}; the quantization
of Lagrangians for gravity, which requires the computations of
scalar counterterms formed by perturbation of the metric and the
Riemann tensor \cite{GoroffSagnotti}; the classification of metrics
using the Cartan-Karlhede algorithm, involving a number (7 at most)
of derivatives of the components of the Riemann tensor in some basis
\cite{PSI2000}. See also the Introduction of Ref. \cite{CBCR02} for a
list of applications of the invariants of the Riemann tensor.

In this article we present for the first time the {\it Invar} tensor
package, a fast manipulator of expressions containing the Riemann
tensor (or its associates Ricci, Weyl, etc.), focusing on the
mentioned subproblem of scalar expressions. Further generalizations
will be reported in the future. The package has been implemented
for both Mathematica and Maple algebraic systems.

\section{The problem}
A metric $g_{ab}$ on a $d$-dimensional manifold has a Riemann
tensor with $d^2(d^2-1)/12$ independent components (assume $d\ge
3$). Of these, $d(d-1)/2$ change under rotations of the frame used
to define the components and hence have no invariant meaning. The
other $N_d\equiv(d+3)d(d-1)(d-2)/12$ components have invariant
meaning and can be combined into an infinite number of scalar
polynomials of the Riemann tensor (the so called {\em invariants}
of the Riemann tensor) \cite{Weinberg}. In general it is possible
to define {\em differential} invariants, in which covariant
derivatives (with respect to the Levi-Civita connection associated
to the metric $g$) of Riemann are considered, but we shall restrict
ourselves to the {\em algebraic} invariants of the Riemann tensor in
this article.

We shall be primarily concerned with invariants formed by products
of $n$ Riemann tensors with all indices contracted (which we will
call {\em invariants}), or products of $n$ Riemann tensors and a
single $\epsilon$ tensor (to be called {\em dual invariants}),
also with all indices contracted. In both cases we shall refer to
$n$ as the {\em degree} of the invariant.

The main question is whether it is possible to find a basis of
invariants in terms of which all other invariants can be
expressed. Ideally, that basis should have a minimal number of
members and those members should have minimal degree. For
dimension 4, Narlikar and Karmarkar \cite{NarlikarKarmarkar} have
given a set of $N_4=14$ invariants such that, for generic
Lorentzian metrics, it could be possible to give any invariant in
terms of the objects of the basis, but possibly involving roots of
high degree. For special metrics the expressions could become
singular. A basis of 17 objects have been given, using spinor
theory, by Carminati and Zakhary \cite{CarminatiZakhary} which is
guaranteed to stay regular for all combinations of the 15 Petrov
and 6 Segre types. Using the properties of $3\times 3$ symmetric
matrices, Sneddon \cite{Sneddon} has constructed a basis of 38
invariants such that any other invariant can be given as a
polynomial in that basis. For none of those three sets it is known
how to generate the expressions (called syzygies) for given
invariants in terms of the corresponding bases, though a partial
solution has been given recently by Carminati and Lim
\cite{CarminatiLim}. Those three sets are highly specialized for
4d Lorentzian metrics, and it seems impossible to generalize them
to other cases.

In this article we develop completely new methods to deal with the
same problem. We shall construct a basis of invariants up to degree 7
in invariants and degree 5 in dual invariants, and implement very
efficient algorithms to express any other invariant within those limits
in terms of that basis. We shall see that our basis contains the 25
invariants of Sneddon's basis with lower degrees than those limits.
Our method is purely tensorial, not based on spinors, and hence the
results are valid (or can be easily generalized) to non-Lorentzian
metrics and other dimensions. They can also be generalized to
differential invariants, or even to expressions with free indices.
The method is purely based on the Riemann tensor, and does not require
its decomposition into Ricci and Weyl parts, a dimensionally dependent
decomposition. However, to facilitate the comparison with the previous
literature, mostly based on that decomposition, we also provide tools
to translate from Weyl invariants, as shown in the appendix. We stress
the fact that the basis is constructed simultaneously with all syzygies.

\section{Algorithms}
A general polynomial expression in the Riemann tensor is first
expanded into a sum of monomial products of tensors, each one
canonicalized independently. Unfortunately, there is no known
efficient (i.e. fast and with polynomic scaling in degree) algorithm
to manipulate in real time products of Riemann tensors.
Previous tensor computer algebra systems use a number of special
rules for restricted cases (c.f. the nearly 40 rules of {\em
MathTensor} \cite{MathTensor} for monomials of degree three or less),
or slow though rather general methods, for example those of {\em Tools
of Tensor Calculus} \cite{TTC} (based on the construction of large
linear systems of equations \cite{IK96}) and {\em cadabra}
\cite{cadabra} (using Young tableaux decompositions). In any case, it is
not possible to go beyond degree 4 in computations taking less than a
minute with these systems.

Here we employ a combined approach which uses different algorithms at
four different steps (called A, B, C, and D), depending on which of
four sets of properties of the Riemann tensor we use:
We shall use efficient permutation-group algorithms to canonicalize a
given monomial in real time, with respect to the permutation symmetries,
into one of several thousand possible canonical monomials (step A).
Then we shall use a database of solutions in which we have previously
stored the canonical form of any of those monomials with respect to
harder symmetries (steps B, C and D), which cannot be handled
efficiently in real time.
The construction of the database with all those canonical forms
is carried out using essentially a brute force approach based on
intensive computer power to solve large systems of linear equations.
At each of the steps B, C and D new equations will be generated,
decreasing the number of independent invariants.

We now give some details and examples of the procedures followed
at each step:

\begin{description}
\item[Step A: Permutation symmetries.]
These are the permutation of indices that maintain the tensor
invariant modulo a minus sign. For the Riemann tensor,
\begin{equation}
\label{Rperms}
R_{bacd} = - R_{abcd}, \qquad R_{cdab} = R_{abcd}
\end{equation}
are examples of permutation symmetries. The Riemann tensor obeys
the cyclic symmetries which are not permutation symmetries, but
multiterm symmetries. The symmetries of the metric tensor $g_{a
b}$ and the totally anti-symmetric tensor $\epsilon_{a b c d}$ can
also be fully described in terms of permutation symmetries.

Computational group theory has the tools that one needs to address
this type of symmetries \cite{Portugal}. For example, the permutation
symmetries of the Riemann tensor are described by a subgroup of the
symmetric group acting on 4 points, because Riemann is a rank-4
tensor. The permutation symmetries of a product of $n$ Riemann tensors
is a permutation group acting on $4n$ points, which takes into account
not only the permutation symmetries of each tensor but also the
permutation of factors. Let us call $S$ the permutation group of the
Riemann invariants of a fixed degree. If there are dummy indices in a
tensor expression, there is a new kind of permutation symmetries
that comes from dummy index relabelling. Those are described by
another permutation group which we call $D$. All equivalent index
configurations of a Riemann invariant of degree $n$ are described
by a double coset of $S$ and $D$ in the symmetric group $S_{4n}$.
A list of canonical invariants is given by a transversal set of
the double cosets of $S$ and $D$ in $S_{4n}$. There is no
efficient way to generate a transversal set. We simply generated
randomly invariants until getting one representative for each
double coset. By using GAP \cite{GAP}, we confirmed that the
transversal set is complete. This strategy must be repeated for
each degree. We succeeded in generating the Riemann invariants up
to degree 7 using a few GBytes of RAM memory.

We addressed two kinds of Riemann invariants: (1) a product of $n$
Riemann tensors, which we denote by $I_{n,r}$ and (2) a product of $n$
Riemann tensors and a 4-index $\epsilon$ tensor, denoted by
$D_{n,r}$, where $n$ is the degree and the index $r$ gives the position
of the invariant in the list sorted according to a predefined order
of permutations. The set $I_{n,r}$ for all values of $r$ for a
fixed $n$ is a canonical transversal set with respect to the
permutation symmetries involved. The same holds for $D_{n,r}$. Those
are the invariants that we need to manipulate for the rest of the
article. The number of invariants per degree is shown in Table
\ref{table::step1}. Note for example for the degree 7 invariants
the enormous decrease from the complete set of $28!\sim 3\cdot
10^{29}$ invariants to 19610 canonical ones.

\begin{table}
\begin{center}
\begin{tabular}{|c|cc|}
\hline
Degree & Invariants & Dual invariants \\ %
\hline
1      & 1         & 1     \\ %
2      & 4         & 5     \\ %
3      & 13        & 35    \\ %
4      & 57        & 288   \\ %
5      & 288       & 3031  \\ %
6      & 2070      &   --  \\ %
7      & 19610     &   --  \\ %
\hline
\end{tabular}
\end{center}
\caption{\label{table::step1}
Number of independent invariants and dual-invariants after imposing
the permutation symmetries.}
\end{table}

As an example, the monomial of degree 6
$$
R^{bca}{}_c R_{gdh}{}^e R^{gfh}{}_f R_{kba}{}^i R^{kmd}{}_l R_{mei}{}^l
$$
is converted after step A into
\begin{equation} \label{I6_246}
I_{6,246} \equiv
R^{ab} R^{cd} R_a{}^e{}_b{}^f R_c{}^g{}_d{}^h R_e{}^i{}_g{}^j R_{fjhi}
\end{equation}

If we remove those invariants which are explicit products of invariants
of smaller degree, then the numbers decrease even more, and are those
shown in columns A of Table \ref{table::numbers}.
\item[Step B: The cyclic symmetry.] This is a multiterm symmetry
involving three instances of index configurations of the Riemann
tensor:
\begin{equation} \label{Rcyclic}
R_{abcd} + R_{acdb} + R_{adbc} = 0 ,
\end{equation}
which expresses the fact that the Riemann tensor has vanishing
totally antisymmetric part; that is, (\ref{Rcyclic}) is equivalent
to $R_{[abcd]}=0$ applying (\ref{Rperms}). Using this symmetry 
on the tensors of each of the canonical invariants obtained
after step A
we generate many new equations relating three invariants of the same
degree, hence reducing the number of independent invariants to
those in columns B of Table \ref{table::numbers}. 
A simple example, applying the cyclic symmetry on the last Riemann
tensor of $I_{6,246}$, defined in (\ref{I6_246}), is
$$ I_{6,246}-I_{6,243}-I_{6,245} = 0, $$
so that the invariant $I_{6,246}$ is not independent after step B.
Note that so far
our results concerning the non-dual invariants are valid for the
Riemann tensor of any metric in any dimension. Those for the dual
invariants are only valid in 4 dimensions.
\item[Step C: Dimensionally dependent identities.] Antisymmetrization
over more than $d$ indices in dimension $d$ gives zero. This
simple fact can be used to generate new relations (Lovelock type
\cite{Edgar}), which will be therefore valid only for dimensions
smaller or equal to $d$. From now on we restrict ourselves to
dimension 4, and hence generate new equations among the invariants
by antisymmetrizing over five indices the independent invariants
that were output by steps A and B.
For example, antisymmetrizing over the contravariant indices
$b,c,e,f,h$ of (\ref{I6_246}) gives a linear
relation among 12 of the 270 independent invariants of degree 6 after
step B.

This step requires solving large systems of linear equations. For
example, for the degree 7 invariants, we generated about 4
equations for each independent invariant, producing a system of
more than 5000 equations with integer coefficients for 1639
unknowns. Though the initial integers and the final rational
numbers had only a few digits, some intermediate integer numbers
had more than 900 digits. One may reduce the number of equations
and still get the correct answer, but the number of equations
cannot be too close to the number of independent invariants (1639
for degree 7).
\item[Step D: Signature dependent identities.] Now we have two
collections of invariants: those containing the totally antisymmetric
tensor (the ``duals'') and those not containing it (the ``non-duals'').
We can relate them by expansion of the product of two $\epsilon$
tensors using the formula
\begin{equation}
\epsilon^{a_1\ldots a_n}\epsilon_{b_1\ldots b_n} = \sigma
\left|
\begin{array}{ccc}
\delta^{a_1}_{b_1} & \ldots & \delta^{a_n}_{b_1} \\ %
\vdots & & \vdots \\ %
\delta^{a_1}_{b_n} & \ldots & \delta^{a_n}_{b_n}
\end{array}
\right|,
\end{equation}
where $\sigma$ represents the sign of the determinant $g$ of the metric.
That means now that the result will depend on that sign. We follow the
convention $\epsilon_{0123}=\sqrt{|g|}$, and hence $\epsilon^{0123}=
\sigma/\sqrt{|g|}$.

This allows us to generate new equations by multiplying pairs of dual
invariants with degrees $n_1$ and $n_2$. This generates a polynomial
in non-dual invariants of total degree $n_1+n_2$. In general this
polynomial contains an invariant of degree $n_1+n_2$ and hence a
high degree invariant can be expressed in terms of the product of two
lower degree dual invariants (the latter play the role of ``square
roots'' of the former). In this way the degree of the elements of the
basis can be reduced (see an application of this in appendix
\ref{appendix::NK} for the Narlikar and Karmarkar basis). However,
this method highly depends on the dimension of the manifold, and on the
signature of the metric (through the sign $\sigma$).
As an example, starting from 
$D_{3,2}\equiv -R^{ab}R^{cd}R_{ac}{}^{ef}\epsilon_{bdef}$
we find
$$
\sigma\,  D_{3,2}^2 = 4 I_{6,11} + 16 I_{6,19} + 4 I_{6,23}.
$$
For definitions of all canonical invariants see the package files.

\end{description}
\begin{table}
\begin{center}
\begin{tabular}{|c|cc|cc|cc|cc|}
\hline
Degree & A     & A$^*$ & B    & B$^*$ & C & C$^*$ & D & D$^*$ \\ %
\hline
1      & 1     & 1     & 1    & 0     & 1 & 0     & 1 & 0     \\ %
2      & 3     & 4     & 2    & 1     & 2 & 1     & 2 & 1     \\ %
3      & 9     & 27    & 5    & 6     & 3 & 2     & 3 & 2     \\ %
4      & 38    & 232   & 15   & 40    & 4 & 1     & 3 & 1     \\ %
5      & 204   & 2582  & 54   & 330   & 5 & 2     & 3 & 2     \\ %
6      & 1613  &--     & 270  &--     & 8 &--     & 4 &--     \\ %
7      & 16532 &--     & 1639 &--     & 7 &--     & 3 &--     \\ %
\hline
\end{tabular}
\end{center}
\caption{\label{table::numbers}
Number of independent invariants or dual-invariants (denoted with a
star) after imposing the different types of relations: A) permutation
symmetries, B) cyclic symmetry, C) dimensionally dependent relations
(for dimension 4), and D) signature dependent relations (for a metric
of negative determinant).}
\end{table}

\section{Implementation}
We have implemented the Invar tensor package on Mathematica and
Maple computer algebra systems. The Mathematica implementation is
on top of the xTensor package \cite{xTensor} and the Maple
implementation on top of the Canon tensor package \cite{Canon}.
The underlying commands for tensor manipulation of these packages
are fully available, such as metric contraction, conversion of
Riemann to Weyl and vice-versa. Both versions share a database
with the syzygies up to degree 7 (degree 5 for duals). The
invariants are stored as permutations in the disjoint cyclic
notation.

The main command of the Invar package, \texttt{RiemannSimplify},
canonicalizes tensor expressions with Riemann invariants up to
degree 7 and dual invariants up to degree 5. The canonicalization
process follows the steps described in Table \ref{table::numbers}.
Each symmetry kind is used one at a time: first the permutation
symmetries, second the cyclic symmetries, third the dimension
dependent identities, and finally signature dependent identities.
There is an option in \texttt{RiemannSimplify} that prevents the
use of the dimension dependent identities, for users that wish to
work with manifolds of generic dimension.

The user has access to all tables of independent invariants and
reducing equations. Besides, there are commands to convert Riemann
invariants from tensor notation to  $I_{n,r}$ and $D_{n,r}$
notation (\texttt{RInv[n,r]} and \texttt{DualRInv[n,r]} in the
package) and vice-versa. It is also possible to relate the
Narlikar and Karmarkar basis with these invariants (see appendix
\ref{appendix::NK}).

Tensors are indexed objects such as \texttt{Riemann[a,-b,-c,-d]},
which stands for $R^a\,_{b c d}$. Contravariant indices are
positive and covariant are negative. For example, the expression
$$
\epsilon^{a\,b\,c\,d}\,R_{a\,b}\,^{e\,f}\,R_{c\,e\,f}\,^{g}\,R_{d}\,^{h\,i\,j}\,
R_{g\,i}\,R_{h\,j}+\frac{(\epsilon^{a\,b\,c\,d}\,R_{a\,b}\,^{e\,f}\,
R_{c\,d\,e\,f})\,(R^{g\,h\,i\,j}\,R_{g\,i}\,R_{h\,j})}{8} $$
is identically zero in dimension 4 as can be verified by employing
the command
\begin{verbatim}
    expr = epsilon[a,b,c,d]*
        R[-a,-b,e,f]*R[-c,-e,-f,g]*R[-d,h,i,j]*R[-g,-i]*R[-h,-j]+
        1/8*(epsilon[a,b,c,d]*R[-a,-b,e,f]*R[-c,-d,-e,-f])*
        (R[g,h,i,j]*R[-g,-i]*R[-h,-j]);
    RiemannSimplify[ expr ]
\end{verbatim}
in the Mathematica implementation of the Invar package, or by
\begin{verbatim}
    RiemannSimplify( expr );
\end{verbatim}
\[ 0 \]%
in the Maple implementation.

The list of independent dual invariants of degree 3 and the
respective tensor expressions can be generated by the following
command in Mathematica
\begin{verbatim}
 Do [ If [ InvSimplify[ DualRInv[3,i] ] == DualRInv[3,i],
           Print[ DualRInv[3,i] -> InvToRiemann[ DualRInv[3,i] ] ]
         ],
      { i, MaxDualIndex[ 3 ] } ]
\end{verbatim}
and in Maple
\begin{verbatim}
 for i to MaxDualIndex( 3 ) do
     if RiemannSimplify( DualRInv[3,i] ) = DualRInv[3,i] then
         print( DualRInv[3,i] = InvToRiemann( DualRInv[3,i] ) )
     end if
 end do;
\end{verbatim}
\[
{\mathit{DualRInv}_{3, 7}}=\mathit{R}\ ^{\mathit{a1}} \
^{\mathit{a2}}{_{\mathit{a1}}}\ ^{\mathit{a3}}\mathit{
R}{_{\mathit{a2}}}\ ^{\mathit{a4}}\ ^{\mathit{a5}} \
^{\mathit{a6}}\mathit{R}{_{\mathit{a4}}}\ ^{
\mathit{a7}}{_{\mathit{a7}}}\ ^{\mathit{a8}}\epsilon
{_{\mathit{a3}}}{_{\mathit{a5}}}{_{\mathit{a6}}}{\ _{\mathit{a8}}}
\]
\[ {\mathit{DualRInv}_{3, 20}}= \\ \mathit{R}\ ^{\mathit{a1}}\
^{\mathit{a2}}\ ^{ \mathit{a3}}\
^{\mathit{a4}}\mathit{R}{_{\mathit{a1 }}}{_{\mathit{a2}}}\
^{\mathit{a5}}\ ^{\mathit{a6}}
\mathit{R}{_{\mathit{a3}}}{_{\mathit{a4}}}\ ^{ \mathit{a7}}\
^{\mathit{a8}}\epsilon {_{\mathit{a5}}}
{_{\mathit{a6}}}{_{\mathit{a7}}}{_{\mathit{a8}}} \]

The complete list of commands is described in the appendix
\ref{appendix::MathematicaCommands} for
the Mathematica implementation and in the appendix
\ref{appendix::MapleCommands} for the Maple implementation.

%

\section{Conclusions}

We have described a systematic procedure to obtain all relations
(called syzygies)
among the scalar invariants of the Riemann tensor. The procedure
relies on a fast algorithm that canonicalizes Riemann monomials
using the permutation symmetries of the Riemann tensor. The cyclic
relations and the dimension and signature dependent identities are
calculated using antisymmetrization methods. The resulting system
of equations is solved and stored in a database using the disjoint
cyclic notation for permutations. Once the database has been produced
and optimized, this approach allows much faster simplifications than
real time methods and requires much less memory. Both approaches use
exponential algorithms, but our method does it only once, and including
the dimensionally-dependent identities (not included in the previously
mentioned real-time systems).
After loading the database, expressions with scalar invariants are
quickly simplified in terms of independent invariants of minimum degree.

The Invar tensor package, implemented both in Mathematica and
Maple on top of the xTensor and Canon tensor packages respectively,
allows the user to manipulate the database of invariants and
syzygies. The database is the same for both implementations.

The method used to build the database can be generalized
straightforwardly to find syzygies among differential invariants
and monomials of the Riemann tensor with free indices. We are
currently analyzing such extensions.

\section*{Acknowledgements} JMM thanks the LNCC for support and
hospitality. JMM was supported by the Spanish MEC under the
research project FIS2005-05736-C03-02. Part of the computations
were performed at the {\em Centro de SuperComputaci\'on de
Galicia} (CESGA).

\appendix

\section{Independent invariants}

The final list of 25 independent invariants is given in table
\ref{indeps4}. The invariants are internally handled in the {\em Invar}
package as permutations of the indices of a product of Riemann tensors
(and a 4-index $\epsilon$ tensor for duals), as given in the
`Only Riemann' column. When the contracted Riemann tensors are
transformed into Ricci the canonicalized expressions give a relative
sign in several cases (`Riemann and Ricci' column).
\begin{table}
$$
\begin{array}{|c|l|l|}
\hline\hline
{\rm Invariant} & {\rm Only\ Riemann} & {\rm Riemann\ and\ Ricci} \\
\hline\hline
I_{1,1} & R^{ab}{}_{ab}
        & R \\
\hline
I_{2,1} & R^{ab}{}_a{}^c R_b{}^d{}_{cd}
        & R^{ab}R_{ab}, \\
I_{2,2} & R^{abcd}R_{abcd}
        & R^{abcd}R_{abcd} \\
D_{2,2} & R^{abcd} R_{ab}{}^{ef} \epsilon_{cdef}
        & R^{abcd} R_{ab}{}^{ef} \epsilon_{cdef} \\
\hline
I_{3,1} & R^{ab}{}_a{}^c R_b{}^d{}_d{}^e R_c{}^f{}_{ef}
        & - R^{ab}R_a{}^cR_{bc}, \\
I_{3,2} & R^{ab}{}_a{}^c R_b{}^d{}_c{}^e R_d{}^f{}_{ef}
        & R^{ab}R^{cd}R_{acbd}, \\
D_{3,2} & R^{ab}{}_a{}^c R_b{}^{def} R_d{}^g{}_g{}^h \epsilon_{cefh}
        & - R^{ab} R^{cd} R_{ac}{}^{ef} \epsilon_{bdef} \\
I_{3,5} & R^{abcd}R_{ab}{}^{ef}R_{cdef} &
          R^{abcd}R_{ab}{}^{ef}R_{cdef} \\
D_{3,13} & R^{abcd} R_{ab}{}^{ef} R_{cd}{}^{gh} \epsilon_{efgh}
         & R^{abcd} R_{ab}{}^{ef} R_{cd}{}^{gh} \epsilon_{efgh} \\
\hline
I_{4,1} & R^{ab}{}_a{}^c R_b{}^d{}_d{}^e R_c{}^f{}_f{}^g R_e{}^h{}_{gh}
        & R^{ab}R_a{}^cR_b{}^dR_{cd}, \\
I_{4,5} & R^{ab}{}_a{}^c R_b{}^{def} R_c{}^g{}_{ef} R_d{}^h{}_{gh}
        & R^{ab} R^{cd} R_{ac}{}^{ef} R_{bdef} \\
I_{4,7} & R^{ab}{}_a{}^c R_b{}^d{}_c{}^e R_d{}^f{}_e{}^g R_f{}^h{}_{gh}
        & R^{ab} R^{cd} R_a{}^e{}_b{}^f R_{cedf} \\
D_{4,7} & R^{ab}{}_a{}^c R_b{}^d{}_c{}^e R_d{}^{fgh} R_f{}^i{}_i{}^j
          \epsilon_{eghj}
        & - R^{ab} R^{cd} R_a{}^e{}_b{}^f R_{ce}{}^{gh} \epsilon_{dfgh}
          \\
\hline
I_{5,2} & R^{ab}{}_a{}^c R_b{}^d{}_c{}^e R_d{}^f{}_f{}^g R_e{}^h{}_h{}^i
          R_g{}^j{}_{ij}
        & R^{ab} R_a{}^c R_b{}^d R^{ef} R_{cedf} \\
D_{5,2} & R^{ab}{}_a{}^c R_b{}^d{}_d{}^e R_c{}^f{}_f{}^g R_e{}^{hij}
          R_h{}^k{}_k{}^l \epsilon_{gijl}
        & - R^{ab} R_a{}^c R_b{}^d R^{ef} R_{ce}{}^{gh} \epsilon_{dfgh}
          \\
I_{5,8} & R^{ab}{}_a{}^c R_b{}^d{}_c{}^e R_d{}^f{}_e{}^g R_f{}^h{}_h{}^i
          R_g{}^j{}_{ij}
        & - R^{ab} R_a{}^c R^{de} R_b{}^f{}_c{}^g R_{dfeg} \\
I_{5,33} & R^{ab}{}_a{}^c R_b{}^d{}_c{}^e R_d{}^{fgh}
           R_e{}^i{}_{gh} R_f{}^j{}_{ij}
         & R^{ab} R^{cd} R_a{}^e{}_b{}^f R_{ce}{}^{gh} R_{dfgh} \\
D_{5,76} & R^{ab}{}_a{}^c R_b{}^d{}_c{}^e R_d{}^f{}_e{}^g R_f{}^{hij}
           R_h{}^k{}_k{}^l \epsilon_{gijl}
         & - R^{ab} R^{cd} R_a{}^e{}_b{}^f R_c{}^{ghi} R_{egf}{}^j
           \epsilon_{dhij} \\
\hline
I_{6,6} & R^{ab}{}_a{}^c R_b{}^d{}_d{}^e R_c{}^f{}_f{}^g R_e{}^{hij}
          R_g{}^k{}_{ij} R_h{}^l{}_{kl}
        & R^{ab} R_a{}^c R_b{}^d R^{ef} R_{ce}{}^{gh} R_{dfgh} \\
I_{6,8} & R^{ab}{}_a{}^c R_b{}^d{}_c{}^e R_d{}^f{}_e{}^g R_f{}^h{}_h{}^i
          R_g{}^j{}_j{}^k R_i{}^l{}_{kl}
        & R^{ab} R_a{}^c R_b{}^d R^{ef} R_c{}^g{}_d{}^h R_{egfh} \\
I_{6,47} & R^{ab}{}_a{}^c R_b{}^d{}_d{}^e R_c{}^f{}_e{}^g R_f{}^{hij}
           R_g{}^k{}_{ij} R_h{}^l{}_{kl}
         & - R^{ab} R_a{}^c R^{de} R_b{}^f{}_c{}^g R_{df}{}^{hi}
           R_{eghi} \\
I_{6,242} & R^{ab}{}_a{}^c R_b{}^d{}_c{}^e R_d{}^{fgh} R_e{}^i{}_{gh}
            R_f{}^j{}_i{}^k R_j{}^l{}_{kl}
          & R^{ab} R^{cd} R_a{}^e{}_b{}^f R_c{}^g{}_d{}^h R_{eg}{}^{ij}
            R_{fhij} \\
\hline
I_{7,14} & R^{ab}{}_a{}^c R_b{}^d{}_d{}^e R_c{}^f{}_e{}^g
           R_f{}^h{}_g{}^i R_h{}^j{}_j{}^k R_i{}^l{}_l{}^m
           R_k{}^n{}_{mn}
         & - R^{ab} R_a{}^c R_b{}^d R^{ef} R_e{}^g R_c{}^h{}_d{}^i
           R_{fhgi} \\
I_{7,55} & R^{ab}{}_a{}^c R_b{}^d{}_d{}^e R_c{}^f{}_f{}^g
           R_e{}^h{}_g{}^i R_h{}^{jkl} R_i{}^m{}_{kl} R_j{}^n{}_{mn}
         & R^{ab} R_a{}^c R_b{}^d R^{ef} R_c{}^g{}_d{}^h R_{eg}{}^{ij}
           R_{fhij} \\
I_{7,391} & R^{ab}{}_a{}^c R_b{}^d{}_c{}^e R_d{}^{fgh} R_e{}^i{}_{gh}
            R_f{}^j{}_i{}^k R_j{}^l{}_l{}^m R_k{}^n{}_{mn}
          & - R^{ab} R_a{}^c R^{de} R_b{}^f{}_c{}^g R_d{}^h{}_e{}^i
            R_{fh}{}^{jk} R_{gijk} \\
\hline\hline
\end{array}
$$
\caption{\label{indeps4}
All independent invariants for dimension 4, up to degrees 7 (non-duals)
and 5 (duals).}
\end{table}

\section{The Narlikar and Karmarkar basis}
\label{appendix::NK}

This basis does not contain dual invariants, which increases the degree
of the elements of the basis from 6 to 7. It is equivalent to the
invariants
$I_{1,1}$, $I_{2,1}$, $I_{2,2}$, $D_{2,2}$, $I_{3,1}$, $I_{3,2}$,
$D_{3,2}$, $I_{3,5}$, $D_{3,13}$, $I_{4,1}$, $I_{4,5}$, $I_{5,2}$,
$D_{5,2}$ and $I_{6,6}$, if $D_{2,2}\ne0$. As shown below, $D_{2,2}$
is a pure Weyl invariant, and hence it does not vanish for non-flat
spacetimes (not even for curved {\it vacuum} spacetimes).

Pure Ricci invariants:
\begin{eqnarray}
&& I_1 \equiv R = I_{1,1}, \\
&& I_2 \equiv R^{ab}R_{ab} = I_{2,1}, \\
&& I_3 \equiv R^{ab}R_a{}^cR_{bc} = - I_{3,1}, \\
&& I_4 \equiv R^{ab}R_a{}^cR_b{}^dR_{cd} = I_{4,1},
\end{eqnarray}
Pure Weyl invariants:
\begin{eqnarray}
J_1 &\equiv & W^{abcd} W_{abcd} =
I_{2,2}-2I_{2,1}+\frac{1}{3}I_{1,1}^2 , \\ J_2 &\equiv &W^{abcd}
W_{ab}{}^{ef}W_{cdef} \nonumber \\ &=&
I_{3,5}-6I_{3,2}+6I_{3,1}-\frac{1}{2}I_{1,1}I_{2,1}+
7I_{1,1}I_{2,1}-\frac{17}{18}I_{1,1}^3 ,\\ J_3 &\equiv &W^{abcd}
W_{ab}{}^{ef}W_{cd}{}^{gh}W_{efgh} -\frac{1}{4} J_1^2 =
\frac{\sigma}{16} D_{2,2}^2 , \\ J_4 &\equiv & W^{abcd} W^{efgh}
W_{abef}W_{cd}{}^{ij}W_{ghij} -\frac{5}{12} J_1 J_2 \nonumber
\\ &=& - \frac{5\sigma}{96} D_{2,2}(-2D_{3,13}-4D_{3,2}+
D_{2,2}I_{1,1}) ,
\end{eqnarray}
Mixed invariants:
\begin{eqnarray}
K_1 &\equiv& R^{ab}R^{cd}W_{acbd} = I_{3,2} - I_{3,1}
-\frac{7}{6}I_{1,1}I_{2,1} +\frac{1}{6}I_{1,1}^3 , \\
K_2 &\equiv&
R^{ab}R^{cd}W_{ac}{}^{ef}W_{bdef}  \nonumber \\&=&
I_{4,5} + 2I_{4,1} -\frac{4}{3}I_{1,1} I_{3,2} - 3 I_{1,1} I_{3,1}
-I_{2,1}^2 + \frac{41}{18} I_{1,1}^2 I_{2,1}
-\frac{5}{18}I_{1,1}^4 , \\ K_3 &\equiv&
R^{ab}R^{cd}W_{ac}{}^{ef}W_{bd}{}^{gh}W_{efgh} -\frac{1}{4} J_1
K_1 +\frac{1}{12} (I_2-I_1^2) J_2 \nonumber \\&=& -
\frac{\sigma}{16} D_{2,2} D_{3,2} ,
\\
K_4 &\equiv& R^{ab}R_a{}^cR^{de}R_d{}^f W_{becf} \nonumber \\&=&
-2I_{5,2} -\frac{7}{6}I_{1,1}I_{4,1}+\frac{1}{2}(I_{2,1}
+I_{1,1}^2)I_{3,2} -\frac{1}{6}(I_{2,1}-11I_{1,1}^2)I_{3,1}
\nonumber \\ && +\frac{2}{3}I_{1,1}I_{2,1}^2
-\frac{4}{3}I_{1,1}^3I_{2,1}+\frac{1}{6}I_{1,1}^5, \\
K_5 &\equiv&
R^{ab}R_a{}^cR^{de}R_d{}^fW_{be}{}^{gh}W_{cfgh} = \nonumber \\&=&
- 2 I_{6,6} - \frac{4}{3} I_{1,1} I_{5,2}
+ \frac{1}{2} (I_{2,1}+I_{1,1}^2) I_{4,5}
+ \frac{1}{4} (2I_{2,1}-I_{2,2})I_{4,1} \nonumber \\ &&
-\frac{13}{18}I_{1,1}^2(I_{4,1}+I_{1,1}I_{3,1})
+\frac{1}{3}(I_{1,1}I_{2,1}-I_{1,1}^3-2I_{3,1})I_{3,2}
+\frac{1}{3}I_{3,1}^2 \nonumber \\ &&
-\frac{1}{6}I_{1,1}I_{2,2}I_{3,1}
+\frac{7}{9}I_{1,1}I_{2,1}I_{3,1}
+\frac{1}{24}(3I_{2,1}^2-I_{1,1}^4)I_{2,2}
-\frac{1}{4}I_{2,1}^3 \nonumber \\ &&
+\frac{7}{18}I_{1,1}^2I_{2,1}^2
+\frac{1}{36}I_{1,1}^4(I_{1,1}^2-8I_{2,1})
, \\
K_6
&\equiv&
R^{ab}R_a{}^cR^{de}R_d{}^fW_{be}{}^{gh}W_{cf}{}^{ij}W_{ghij} -
\frac{1}{4} J_1 K_4 +\frac{1}{12}(I_4-I_2^2) J_2 \nonumber \\ &=&
\frac{\sigma}{32}D_{2,2}(4D_{5,2}-D_{3,2}(I_{1,1}^2+I_{2,1})).
\end{eqnarray}

\section{Main commands in the Mathematica implementation}
\label{appendix::MathematicaCommands}

This appendix lists the main commands of the Invar tensor package
in the Mathematica implementation version. This version is written
on top of the xTensor package \cite{xTensor}.

\texttt{InvSimplify[expr, sl]} is the general simplifier of scalar
invariants in \texttt{expr} when expressed with heads \texttt{RInv} and
\texttt{DualRInv}. The optional argument \texttt{sl} is an integer
specifying the simplification level: 1 (only permutation symmetries),
2 (also cyclic symmetries), 3 (also dimensionally-dependent
identities) and 4 (also signature-dependent identities), which
correspond to columns A, B, C and D of table \ref{table::numbers},
respectively. Its default value is 4.

\texttt{RiemannToInv[expr]} converts the algebraic expression
\texttt{expr} containing Riemann and Ricci tensors into an
expression with objects with head \texttt{RInv} and \texttt{DualRInv}.
Conversely, \texttt{InvToRiemann[expr]} transforms those objects in
\texttt{expr} into their corresponding Riemann expressions.

\texttt{RiemannSimplify[expr]} is equivalent to the consecutive action
on \texttt{expr} of \texttt{RiemannToInv}, then \texttt{InvSimplify},
and then \texttt{InvToRiemann}.

\texttt{Invs[sl, d]} gives a list off all independent invariants of
degree \texttt{d} after simplification level \texttt{sl}.
\texttt{DualInvs[sl, d]} does the same for dual invariants.

\texttt{MaxIndex[d]} and \texttt{MaxDualIndex[d]} return
the maximum value that the index \texttt{n} can assume in
\texttt{RInv[d, n]} and \texttt{DualRInv[d, n]}
respectively.

These are functions of xTensor which can be useful in the Invar
package:

\texttt{DefManifold[M, dim, \{a, b, c, ...\}, s]} defines a manifold
\texttt{M} of dimension \texttt{dim} and associates the abstract
indices \texttt{\{a, b, c, ..., s\}} to it.

\texttt{DefMetric[$\sigma$, g[-a, -b], CD, \{";", "$\nabla$"\}]}
defines the metric \texttt{g} on the manifold \texttt{M}, such that
its determinant has sign \texttt{$\sigma$}. Its Levi-Civita
connection \texttt{CD} is also defined, as well as the associated
curvature tensors \texttt{RiemannCD[-a,-b,-c,d]},
\texttt{RicciCD[-a,-b]}, \texttt{RicciScalarCD[]} and
\texttt{WeylCD[-a,-b,-c,-d]}, and the totally
antisymmetric tensor \texttt{epsilong[-a,-b,-c,-d]} (in dimension 4).

\texttt{ContractMetric[expr, g]} simplifies contractions with the
metric \texttt{g} in the expression \texttt{expr}.

\texttt{RiemannToWeyl[expr]} expands Riemann tensors into Weyl,
Ricci and RicciScalar parts. \texttt{WeylToRiemann[expr]} performs
the opposite task.

\texttt{RiemannToRicci[expr]} converts contracted Riemann tensors into
Ricci tensors, and contracted Ricci into RicciScalar.

\section{Main commands in the Maple implementation}
\label{appendix::MapleCommands}

This appendix lists the main commands of the Invar tensor package
in the Maple implementation version. This version is written on
top of the Canon package \cite{Canon}. The arguments between
square brackets are optional.

\texttt{RiemannSimplify(expr,[opt])} is a general simplifier of
scalar invariants. It contracts any metric tensor in
\texttt{expr}, simplifies contractions of the Weyl tensor,
converts products of Riemann tensors into RInvs and DualRInvs
whenever possible, reduces them using cyclic and dimension
dependent identities, and simplifies contractions the Riemann
tensor into Ricci and Ricci scalars. The option \texttt{opt}
controls the simplifying level. The default is \texttt{DDI}, which
means that the dimension dependent identities are employed. The
option \texttt{Cyclic} tells \texttt{RiemannSimplify} to use only
the cyclic identities.

\texttt{RiemannToInv(expr)} converts the algebraic expression
\texttt{expr} that has Riemann tensors into an expression with
RInvs and DualRInvs. \texttt{InvToRiemann( expr)} converts the
algebraic expression \texttt{expr} that has RInvs and DualRInvs
into an expression with Riemann tensors.

\texttt{RiemannToRicci(expr)} simplifies contractions of Riemann
tensors into Ricci tensors or Ricci scalars.
\texttt{RicciToRiemann} converts Ricci tensors into Riemann
tensors.

\texttt{WeylToRiemann(expr)} converts Weyl tensors into Riemann
tensors. \texttt{Riemann\ ToWeyl(expr)} converts Riemann tensors
into Weyl tensors.

\texttt{MaxIndex(degree)} and \texttt{MaxDualIndex(degree)} return
the maximum value that the index \texttt{n} can assume in
\texttt{RInv[degree,n]} and \texttt{DualRInv[degree,n]}
respectively.

It follows a short description of commands of the Canon package
useful in the context.

\texttt{TensorDefine(T, n, S, [B])} defines the symmetries of a
tensor. \texttt{T} is the tensor name, \texttt{n} is the number of
indices, \texttt{S} is a set of permutations, and \texttt{B} is
the base. For example, the tensor A with 3 indices and
antisymmetric in the last two indices is defined in the following
way: \texttt{TensorDefine(A,3,{[-1,[[2,3]]]},\ [1,2,3])}. Totally
symmetric or antisymmetric tensors can be defined by using the
option Symmetric or AntiSymmetric as the third argument. The list
of built-in tensors is: Riemann tensor \texttt{R[a,b,c,d]}, Ricci
tensor \texttt{R[a,b]}, Ricci scalar \texttt{R[ ]}, totally
antisymmetric tensor \texttt{epsilon[a,b,c,d]},  metric tensor
\texttt{g[a,b]},  Weyl tensor \texttt{C[a,b,c,d]}.

\texttt{Canonical(expr)} uses the permutation symmetries of the
built-in tensors (or tensors defined by the user) to canonicalize
the tensor expression \texttt{expr}. No further simplification is
performed. Most commands of the package uses Canonical internally.

\texttt{AbsorbMetric(expr)} simplifies contractions with the
metric tensor.

\end{document}